# Digital microfluidics – How magnetically driven orientation of pillars influences droplet positioning

Blandine Bolteau[1, 2, 3], Frédéric Gelebart[1], Jérémie Teisseire[3], Etienne Barthel*[2], Jérôme Fresnais*[1]

[1] Sorbonne Université, CNRS, Laboratoire de Physico-chimie des Electrolytes et Nanosystèmes Interfaciaux,
PHENIX - UMR 8234, F-75252 Paris cedex 05, France.
[2] Sciences et Ingénierie de la Matière Molle, ESPCI Paris, Université PSL, CNRS, Sorbonne Université, 75005 Paris, France
[3] Surface du Verre et Interfaces, UMR 125 CNRS/Saint-Gobain, 39 Quai Lucien Lefranc, F-93303 Aubervilliers, Cedex, France

\* Corresponding authors: jerome.fresnais@sorbonne-universite.fr ; etienne.barthel@espci.fr

# Abstract

Interfaces between a water droplet and a network of pillars produce eventually superhydrophobic, self-cleaning properties. Considering the surface fraction of the surface in interaction with water, it is possible to tune precisely the contact angle hysteresis (CAH) to low values, which is at the origin of the poor adhesion of water droplets, inducing their high mobility on such a surface. However, if one wants to move and position a droplet, the lower the CAH, the less precise will be the positioning on the surface. While rigid surfaces limit the possibilities of actuation, smart surfaces have been devised with which a stimulus can be used to trigger the displacement of a droplet. Light, electron beam, mechanical stimulation like vibration, or magnetism can be used to induce a displacement of droplets on surfaces and transfer them from one position to the targeted one. Among these methods, only few are reversible, leading to anisotropy-controlled orientation of the structured interface with water. Magnetically driven superhydrophobic surfaces are the most promising reprogramming surfaces that can lead to the control of wettability and droplet guidance.

The use of rigid regular pillar arrays has permitted normalizing and understanding the interaction between the contact line of the liquid droplet and the pillars, and more generally the wetting behavior in either the Cassie or the Wenzel states. When elastic and magnetic pillar arrays are considered, new behaviors are expected from the solid water interface. In this work, we synthesized superhydrophobic magnetic elastomer pillar arrays to highlight the role of mechanical stress on the contact line, and its implication on the displacement of a droplet on such a surface. We demonstrated that, controlling the surface structure, the local receding contact angle at the scale of a pillar directed the anchoring of the droplet, leading to the macroscopic control of its displacement on a tilted surface. This comprehensive study gives more insight into the relation between local dewetting and macroscopic movement of the contact line.

## Background and Significance

Superhydrophobic surfaces are produced by the combination of hydrophobic coatings and adapted roughness [1]. They have numerous potential applications that have not been completely realized due to the fragility due to the micro to nanostructures present at the interface. Despite this limitation, a lot of publications have appeared on those materials with generic surface behavior. A huge effort has been made during the last 10 years to understand the Cassie wetting state and its transition toward Wenzel state, especially by exploring the wetting behavior of model surfaces made of pillar arrays [2–4]. Their surface density, pillar height, or individual size impact drastically the stability of the composite wetting (with both material and air) [5]. The Cassie to Wenzel transition is also impacted by the geometry of the pillars, which was demonstrated to be directly dependent on the balance between the elasticity of the contact line and the geometrical distribution of the pillars [6].

Precise positioning of water droplets onto non-wetting surfaces has been demonstrated by the addition of magnetic deformation of the surface [7], distribution of more hydrophilic spots [8], but in majority, vibration or mechanical movement (deformation) of the surface is used to trigger the droplet movement [9,10]. During this macroscopic movement, the contact angle varies drastically, in addition to the pressure of the droplet on the pillar array. This could induce irreversible pinning of the droplet if pillar aspect ratio is not optimized.

Hopefully, flexible surfaces can help to avoid this wetting transition, by either deforming under an external stimulus or absorbing partially the force of the droplet impact [11]. Although natural elastic filament arrays have demonstrated their efficiency in nature [12,13], synthetic ones are more difficult to prepare, due to the high aspect ratio needed to reproduce the hairs observed onto natural surfaces. Fewer are the surfaces that can impose forces on the contact line, for instance by adapting the wetting anisotropy [14]. In this framework, recent publications demonstrated that magnetic elastomer pillar arrays can be used to conveniently adapt the surface topology toward specific and transient wetting behavior. As an example, Seo et al. demonstrated the ability of elastomer superhydrophobic surfaces to guide and merge droplets [15]. Yang et al. prepared magnetic pillar arrays to alternatively control the adhesion of droplets on the magnetically actionable surfaces [16]. Magnetic rigid pillar arrays have been advantageously used to control the wetting and rebound of droplets on demand [17]. Models have been proposed for rigid pillar arrays, such as the evaluation of the contact line dynamic during evaporation by Gauthier et al. [18,19]. They proposed that the contact line is attached to a pillar summit and moves progressively from each configuration during pillar depinning. Moreover, they evidenced the impact of the receding contact angle on the dewetting on rigid pillar arrays and its relevant value compared to its transient values. These evaluations are clearly lacking for deformable pillar arrays.

In this article, we propose, in the first part of this article, a comprehensive study of the deflection of pillar triggered by a magnetic activation placed below the surface. In this configuration, the magnetic gradient generated by the magnets is the driving force for their deflection. More importantly, using several surface fractions of pillars and aspect ratio, we highlight the preponderant role of the dipolar magnetic interaction between pillars on their deflections, which has never been discussed yet. In the second part of the article, we have selected the most effective pillar array that exhibit the largest deflexion. These pillars are used to monitor the dynamic of the contact line during magnetic actuation of pillars, during the evaporation of a droplet or for its movement when the surface is tilted. From local measurement to macroscopic observation, we were able to confirm that the orientation of the pillars modifies temporary the local contact angle (either toward receding or advancing one), which affect the dynamic of the contact line, thus the positioning of the droplet.

## Methods

**Magnetic elastomer**

Magnetic PDMS was prepared with iron microparticles ranging from 0.1 µm to 5 µm in diameter. Particle powder was added to Sylgard part A and mixed with a deflocking rotating mixer for 30 minutes. The amount of microparticles in PDMS was possible up to 80% by weight, that corresponds to 10.2% in volume fractions.

In order to prepare magnetic pillar arrays, soft lithography methods commonly presented in the literature were used [20]. This consisted in preparing pillar arrays using appropriate SU-8 resin layers with targeted thicknesses followed by a UV insolation through a mask. Then, a PDMS (SYLGARD 184) counter-mold is produced from the resin microstructures, silanized to become hydrophobic and avoid adhesion, and used to reproduce the native pillar arrays copies with magnetic PDMS elastomers. More than 25 structures were studied in this work, differing from their surface fraction ($\phi_S$) and the aspect ratio of the pillars (AR). The magnetic elastomer was characterized by Vibrating Sample Magnetometry to evaluate the magnetization at saturation of the as-prepared magnetic PDMS.

**Pattern fabrication**

Superhydrophobic surfaces were prepared by a soft lithography process using an SU-8 photosensitive resin mold. It consists of a square network of cylindrical pillars with diameters ranging from 8.5 to 25.5 µm, and lengths ranging from 31.8 to 103 µm. the mean distance between pillars were tuned to 30, 50, and 70 µm.

For the second part of the study, a pillar array with pillar diameter of 17.1 µm, length 61.6 µm high and the spacing of 50 µm between the center of pillars was used. For this specific pillar array, the SU-8 resin was deposited through spin coating (30 s at 500 rpm, followed by 30 s at 1900 rpm) and preheated at 65 °C for 3 minutes and 95 °C during 9 minutes. The exposure across a transparent mask was achieved with 8 cycles of 30 seconds on and 20 seconds off at 70% of the maximum power of the lamp at 365 nm in a UV-KUBE2. Post-bakes at 65 °C for 2 minutes followed by one at 95 °C for 7 minutes are achieved to ensure a complete polymerization of the resin. Development is conducted with SU8-developper (Propylene Glycol Monomethyl Ether Acetate, PG-MEA). At last, a hard bake is achieved at 150 °C for 15 minutes.

Then, SYLGARD 184 was used to prepare a negative mold of the pillars. After activation with an O2 plasma treatment and silanization with trichloro (1H,1H,2H,2 H-perfluorooctyl) silane in the vapor phase, this mold was employed to reproduce the pillars network using SYLGARD 184. This mold can be exploited several times if mandatory. Those surfaces were characterized by Scanning Electron Microscopy (SEM) and optical microscopy (OLYMPUS BX5TRF-1).

Table 1: characteristics of the different pillar array geometries synthesized in this study. $\lambda$ corresponds to the distance between two pillar centers; $\Phi_s$ corresponds to the surface fraction; AR is the aspect ratio of the pillars, defined as their height (h) divided by their diameter (d).

| Geometry | $\lambda$(µm) | $\Phi_s$ (%) | AR | d (µm) | Standard deviation | h (µm) | Standard deviation |
|---|---|---|---|---|---|---|---|
| A | 50 | 15 | 1.5 | 21.9 | 0.7 | 33 | 0.4 |
| C | 50 | 9.4 | 3.6 | 17.3 | 1.1 | 62.8 | 1.2 |
| D | 50 | 16.9 | 3.1 | 23.2 | 1.4 | 71.7 | 0.8 |
| E | 50 | 12.8 | 3.4 | 20.2 | 1 | 67.9 | 0.9 |
| F | 50 | 2.3 | 3.8 | 8.5 | 0.4 | 31.8 | 0.3 |
| G | 30 | 24 | 3.1 | 16.6 | 0.4 | 50.6 | 1.2 |
| H | 50 | 7.4 | 3 | 15.4 | 0.3 | 46.4 | 0.6 |
| I | 70 | 3.4 | 3.3 | 14.5 | 0.3 | 47.1 | 1.1 |
| J | 50 | 18.3 | 3.4 | 24.3 | 0.5 | 83.4 | 1.3 |
| K | 70 | 6 | 5.3 | 19.4 | 0.7 | 103 | 1.4 |
| L | 50 | 20.4 | 3.7 | 25.5 | 0.4 | 94.7 | 0.7 |
| M | 50 | 17.6 | 3.7 | 23.7 | 0.9 | 86.4 | 1.2 |
| N | 70 | 6.3 | 4.5 | 19.8 | 0.5 | 89.6 | 0.5 |
| O | 30 | 25.5 | 4.9 | 17.1 | 0.4 | 83.8 | 0.4 |
| P | 30 | 22.2 | 4.9 | 15.9 | 0.6 | 78.7 | 0.6 |
| Q | 50 | 11.1 | 4.8 | 18.8 | 0.9 | 89.7 | 1 |
| R | 30 | 33.5 | 3.7 | 19.6 | 0.6 | 72.2 | 0.6 |

The list of all pillar array molds prepared in this work is reported in table 1, where $\lambda$ corresponds to the distance between two pillar centers; $\phi_S$ corresponds to the surface fraction; AR is the aspect ratio of the pillars, defined as their height (h) divided by their diameter (d). The surface fractions range from 2.3% up to 33.5%. The aspect ratios of the pillars range from 1.5 to 5.3.

**Contact angle measurements**

The contact angles (either advancing and receding) were registered using distilled water droplets from 5 µL to 20 µL. The ImageJ software with the Snake plugin was used to measure the contact angles of droplets deposited on the surfaces [21].

**Application of the magnetic field**

To change the orientation of the pillars through magnetic actuation, magnets were attached onto an oscillating motor below the pillar array, that can move to a rotation angle of 180° max with a velocity from 0.01 cm/s up to 4 cm/s. The motor is driven with an Arduino type set-up, and the movement program is prepared with the use of the Arduino IDE software.

Magnets consist of a column of three cubic magnets with 1 cm in size made of Neodynium-iron-Bore magnets (SUPERMAGNETE). The magnets can be approached from 20 mm down to 2 mm from the bottom of the droplet (contact line).

Magnetic field amplitude of the magnets was measured from the surface of the magnets by using a Hall probe. The magnetic field was used to determine the magnetization of the iron microparticles in the elastomer, depending on their concentration.

# Results

## Properties of the magnetic patterned surfaces

Arrays of pillars were synthesized using well-known photolithography techniques adapted for the high aspect ratio pillars. An example of the pillar visualization is given in figure 1. It shows images obtained by SEM of one of the different structures obtained (namely C-type surface). On image figure 1-b, it is possible to distinguish the magnetic microparticles in the pillars as bright spots. A size analysis allows to determine their mean diameter to 550 nm, that evidenced the quite good dispersion of the particles in the pillars.

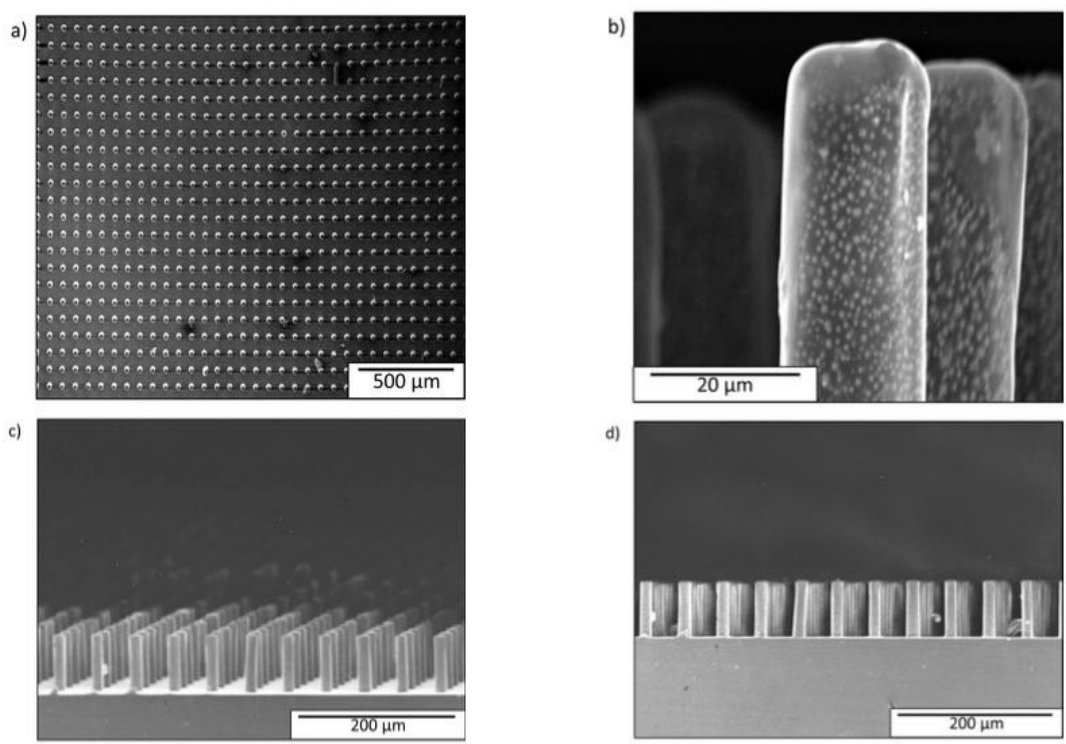

Figure 1: SEM images of pillars array C. Top view (a); side views (c-d); detail of a pillar (b) with magnetic iron particles.

## Wetting

The soft pillar arrays, prepared with various pillar aspect ratios and surface fractions with SYLGARD 184, were characterized by their advancing and receding contact angles, measured by increasing and decreasing the volume of a water droplet, respectively. The advancing contact angles are constant (around 170°) whatever the surface fraction. On the contrary, the receding contact angles decrease with the surface fraction. This is in good agreement with similar works described by Gauthier et al.[18] on rigid pillar array, where the fit of the data does not pass through zero because of a dissipation process due to pinning of the triple line by defect (figure 2). From these measurements, we can confirm that elastic pillar arrays behave as rigid ones, showing that the same theoretical analogies, used for rigid pillar arrays, can be applied to these model surfaces. The classical way to represent the contact angle variation upon the surface fraction is to use the Cassie Baxter model:

$$1 + \cos(\theta_R) = \emptyset_S(1 + \cos(\theta_{R0})), \qquad (1)$$

where $\theta_R$ corresponds to the receding contact angle and $\theta_{R0}$ to the receding contact angle on flat PDMS surface. On figure 2-left, one can see that the data do not cross the origin.

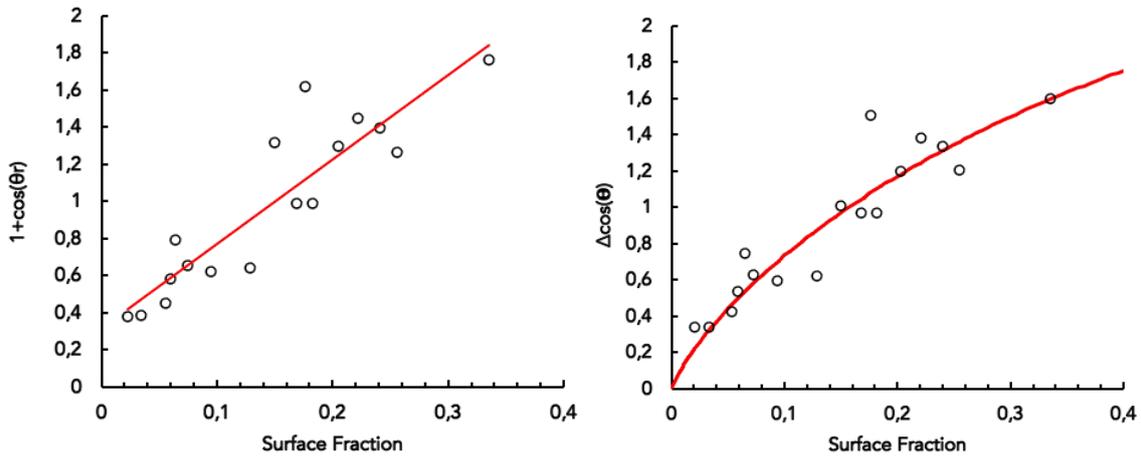

Figure 2: left: variation of surface energy at receding contact angle versus surface fraction in the Cassie regime. The red line is a linear fit of the whole dataset; right: contact angle hysteresis versus surface fraction. The red line corresponds to equation 3 with a=8.5.

This has been described either by by Rivetti et al. [22] and by Reyssat and Quéré [23]. In this later study, they considered the variation of contact angle hysteresis to vary with the surface fraction corrected by a term due to the distortion of the contact line at the scale of a pillar, following equation 2. In their case, they found a value for a of 3.8. The large value obtain in the present study (a=8.5) is in good agreement with the fact that the pillars are larger and not perfectly cylindrical, due to the use of the low-resolved masks used at the photolithography step. The dilute regime of pillars described by Reyssat and Quéré is also confirmed here while the data followed the main trend of the equation 2.

$$\Delta \cos(\theta) \approx \frac{a}{4} \phi \ln\left(\frac{\pi}{\phi}\right), \qquad (2)$$

**Magnetic characterizations**

Each pillar geometry has been used to prepare magnetic pillar arrays based on iron microparticles dispersed in SYLGARD184, part A at 0-10.2% by volume. Magnetic PDMS was characterized magnetically by obtaining the magnetization curves from our home-made vibrating sample magnetometry (VSM) apparatus. The saturation of the magnetization of the particles is obtained above 0.6 T, with a magnetization that is proportional to the volume fraction of particles in the cross-linked PDMS. No magnetic hysteresis was observed at any concentration, showing that the magnetic PDMS remains paramagnetic at room temperature (figure 3).

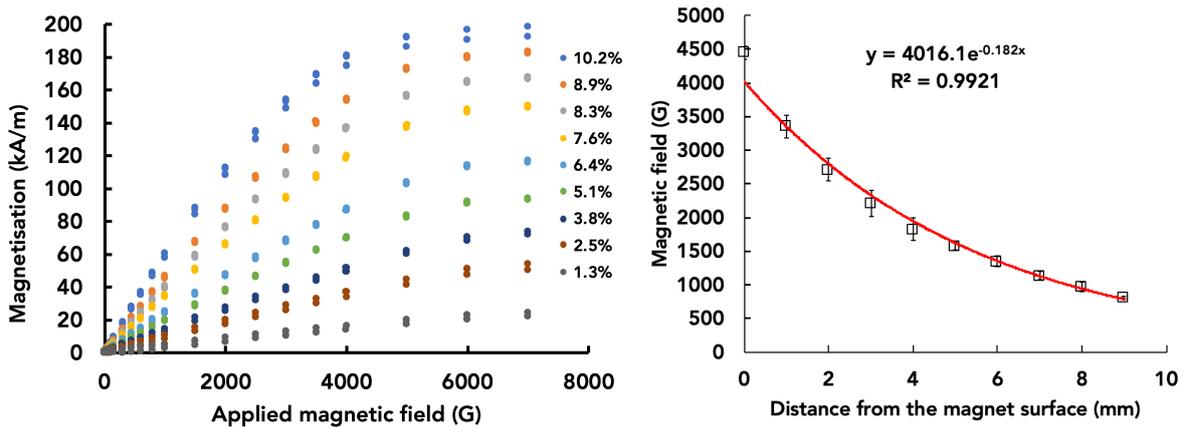

Figure 3: Magnetization curves of magnetic PDMS [left] and magnetic field of the magnets column versus the distance [right].

To actuate the pillars, a column of three cubic magnets (1 cm) is used. The magnetic field provided by this column decreases strongly with the distance to its upper surface. At a distance of 2 mm, the magnetic field amplitude is 27 mT. This magnetic moment is not large enough to completely reach the saturation, thus it is necessary to take the partial magnetization into account in the calculation of the magnetic force applied by the magnet. From this magnetic field amplitude, it is possible to evaluate the saturation of the magnetic PDMS on the magnetization curves. For a volume fraction of particles of 1.27%, the magnetic PDMS reached 43.8% of the saturation value. When the volume fraction reached 10.2%, the magnetization reached 71.3% of the field at saturation. This magnetization will be taken into account in the calculation of the magnetic force.

**Magnetic actuation of the soft pillars**

The pillars are composed partly of paramagnetic iron particles that are magnetized in the presence of the magnets. At the same time, the magnetic gradient increases, allowing the pillar to deflect. The amplitude of deflexion is thus driven by the response of the magnetic material. When the magnets column is displaced below the surface, the pillars are progressively deflected along the magnet trajectory depending on the magnetic force applied. This magnetic force is maximum at the region of highest magnetic gradient, which corresponds to the corner of the magnet cube.

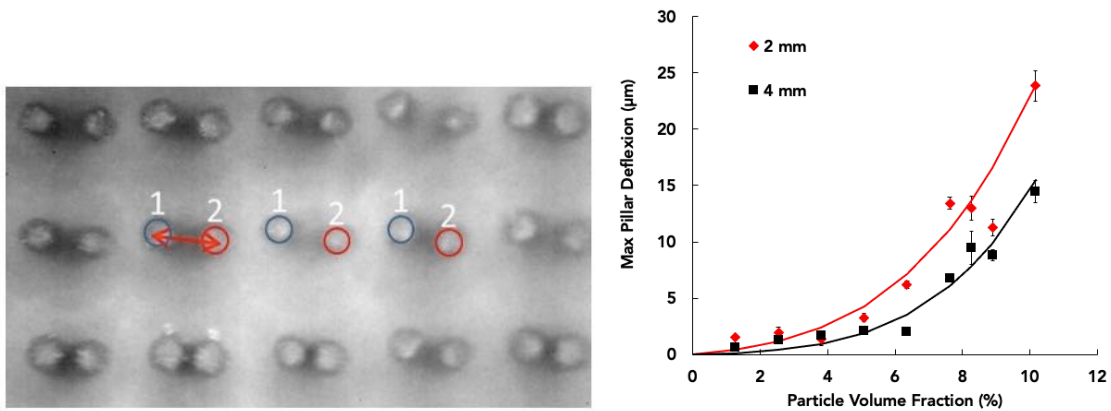

Figure 4: snapshot of the superimposed positions of the pillar at the two extreme positions [surface C; 10.2% of particles]; Deflection versus the volume fraction at two distances between the pillars and the magnet [red: 2 mm; black: 4 mm].

Figure 4 shows a superimposed snapshot of the pillar in the two-extreme deformation [right and left], for pillars C and a volume fraction of particles of 10.2%, allowing the determination of the amplitude of bending of the pillars. It can be seen that all the pillars do not bend with the same amplitude, which is due either to small variations of the diameter, to the defect of the transcription of the mask during the UV insolation process, or to the variability of magnetic content inside the pillars. The maximal deflection increases with the volume fraction of particles, and also as the distance between the magnet and the pillars is decreased (figure 4, right).

**Comparison of magnetic and elastic forces**

From the magnetic characterization, it is possible to determine the magnetic force versus the magnetic loading. It is sometimes difficult to extract the origin of the magnetic actuation that is often described in publications as the result of the magnetic torque between the direction of the magnetic field and the one of the pillars [24]. In our configuration, we strongly hypothesized that the deflexion of the pillars is due to the magnetic gradient rather than the magnetic torque (magnetic gradient rather that magnetic field direction). Of course, the intensity of the magnetic field is crucial for the pillars to acquire a magnetization, but the coupling between the magnetic gradient and the pillar seems the more relevant form of interaction. To prove this, we have calculated the magnetic force, and compared it to the elastic one. Indeed, the magnetic force is expressed as the following:

$$F = m \frac{\Delta \chi}{\rho \mu_0} B \, gradB \quad (3),$$

where $\Delta \chi$ corresponds to the magnetic susceptibility of the magnetic elastomer, m is the mass of magnetic materials, $\rho$ is the density of the elastic magnetic material, $\mu_0$ is the permeability constant, B and gradB are the magnetic field and magnetic gradient amplitude, respectively. $\Delta \chi$ is determined from the magnetization values at the applied magnetic field. The magnetic force was calculated for pillars C, with increasing magnetic particle loading, and at a fixed magnet position relative to the pillars, which highlight the maximum amplitude of deflection (this means that B.gradB is a constant for all the formulations from 0 to 10.2% of particles).

These data were compared to the elastic force obtained by measuring the deflection of pillars with different magnetic particle content. From the optical microscopy images, it is possible to obtain the deflection due to the balance between the magnetic force and the elastic resistance. The elastic force is thus expressed as the following:

$$F_{el} = \frac{3\pi E d^4}{64 h^3} \delta x \quad , \quad (4)$$

where E is the Young modulus, d is the diameter of the pillar, h is the height of the pillar, and $\delta x$ is the deflection measured at the top of the pillars. The calculation of the elastic force was achieved for pillars in C configuration with increasing magnetic loading.

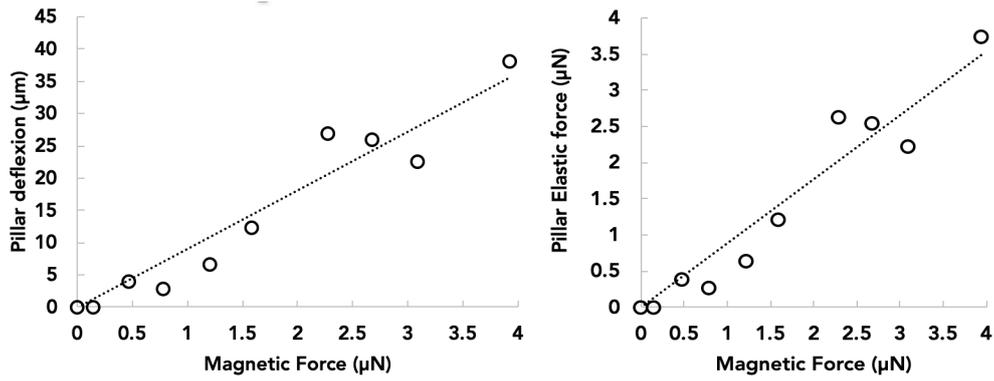

Figure 5: Left: figure 4-b plotted versus the magnetic force (at 2 mm). Right: comparison between magnetic force (equation 3) and elastic force (equation 4).

Figure 5 represents the variation of the force exerted on the pillars versus the magnetic force (5-left) and a comparison between both elastic and magnetic forces determined independently (5-right). Comparing both forces give a slope of 0.89, closed to unity, evidencing the magnetic gradient to be at the origin of the force exerted on the pillars.

Thus, the deflection of the pillar can be classically tuned by the volume fraction of particles in the PDMS hybrid material, as well as by the distance of the column of magnets from the pillars (magnetic gradient) as shown previously.

**Optimization of the magnetic actuation**

We have examined the role of the aspect ratio and the surface fraction of pillars functionalized with 10.2% of particles only. Results are plotted in Fig. 6 left. It is expected that the bending of the pillars varies with their aspect ratio, as demonstrated by Wei et al. [25]. Classically, the deflection, defined as the balance between the magnetic force and the elastic force ($F_{el}$, equation 2), is expected to be proportional to the aspect ratio to the power 4 (red curve in figure 6-right). While almost all pillars followed this predicting law, two samples showed an unexpected result: surface C and E demonstrated significantly larger deflection values than the other samples.

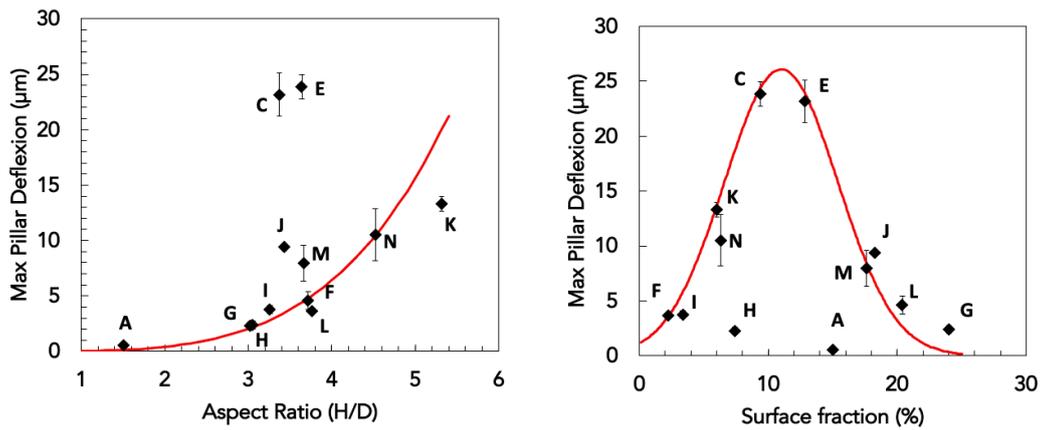

Figure 6: maximum pillar deflection versus aspect ratio (left) of the surface fraction (right) (at 2 mm).

When the same deflection values are plotted versus the surface fraction of pillars (Fig. 6 right), samples C and E can be included in a normal distribution that fits well most of the rest of the data (except H and A), showing that the surface density of the pillars has an important role in their ability to be bent. The reason for this deflection variation remains in the contribution of the magnetic dipolar interactions between pillars that cannot be avoided when considering their interaction with the magnet. Indeed, for low surface fraction values, the pillars seem to follow the model of individual pillars. When the surface fraction increases, the dipolar interactions produce constructive effect, leading to an increase of the deflection compared to expected models (surfaces C and E). At larger surface fractions, the dipolar interactions decrease the deflection amplitude, due to strong pillar-pillar repulsive interactions. For the largest surface fractions, we even observed chaining of pillars at the surface, demonstrating directly that the magnetization of the pillars induces dipolar interactions (SI-11).

To highlight the role of both the aspect ratio and the surface fraction of pillars, it is possible to rationalize the representation of the pillar deflection by normalizing it with the expected one (proportional to $(h/d)^4$), plotted versus the surface fraction (Fig. 7). With this scaling, all the pillars are reasonably accounted for.

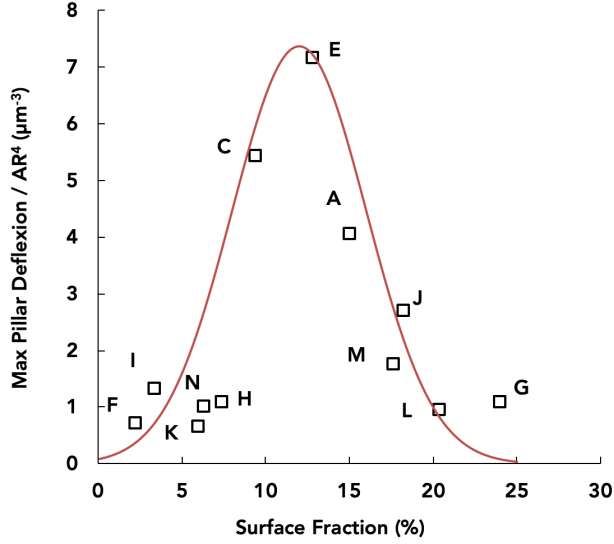

Figure 7: maximum pillar deflection, normalized by $AR^4$, versus surface fraction (at a distance between magnet and pillars of 2 mm).

To understand better why pillar arrays with larger surface area denote lower deflection, we emphasize the impact on magnetic dipolar interaction between pillars due to magnetic polarization under the magnetic field. The classical model of magnetic interaction between two magnetic rods follows the equation 3 [26]:

$$F_{interaction} = -\frac{3\mu_0 m^2}{2\pi L^4} \qquad (5)$$

where m is the magnetic moment of the pillar, L the distance between two adjacent pillars. The magnetic moment is proportional to the volume of the pillar (radius r and length h). this leads to the following equation:

$$F_{interaction} \sim -h^2 \left(\frac{r}{L}\right)^4 \qquad (6)$$

This equation is valid for distance L larger than the pillar diameter, which is fulfilled in our samples where the pillar diameters are typically 20 µm. The deflection of the pillars from different surfaces is plotted versus r/L (figure 8). For large r/L values, the deflection decreases as the dipolar interactions increase, with a slope corresponding to a power law with a coefficient of -4, in good agreement with equation (6). This indicates clearly that the interactions between pillars limit their movement due to detrimental dipolar magnetic interactions.

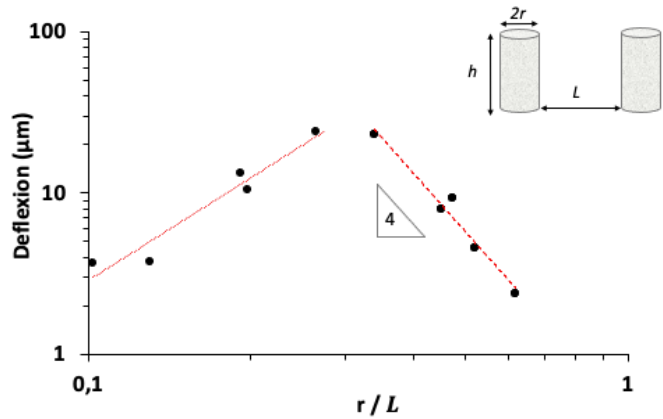

Figure 8: maximum pillar deflection, versus r/L (at a distance between magnet and pillars of 2 mm). The red dotted line corresponds to a power law 2, and the red dash curve is a -4 power law describing magnetic dipolar interaction between pillars. The scheme represents the dimensions of pillars with their respective distance.

At low r/L values, we could expect the dipolar interactions to be neglectable, but it is also important to notice that the distance between pillars seems to have an impact on the deflection, but not with the same variation amplitude. Indeed, the data for low values of r/L can be fitted with a power law with a coefficient of 2. Globally, this means that the interactions, either constructive or detrimental, must be taken into account to have a better control of the mobility of pillar arrays, within this configuration (magnets moving bellow the pillars).

This analysis highlights the fact that, considering the magnetic actuation used in this study, preparation of a magnetic pillar array must be optimized to balance magnetic interaction between pillars that can limit their deflection. This must be optimized by the needs of surface interaction with the fluids. Indeed, decreasing surface fraction (increasing L) reduces the anchoring points of the contact line, leading to more unstable droplets on the surfaces and apparent lower deflection amplitude. When more adhesion is needed, larger surface area could lead to a limitation of the deflexion.

**Static and dynamic depinning of droplets by magnetically actuated soft surfaces**
In the following part of the article, we will only consider the pillars C, with 10.2% of magnetic particles inserted in the PDMS. Indeed, this pillar array demonstrates the largest deflexion values.

**Droplet roll-off and tilt angle asymmetry**

On the surface C, we have considered the sliding of droplets in the so-called Cassie state. We measured the tilting angle at which the droplet of a fixed volume (10 µL) rolls off out of the surface, with static orientation of the pillar against or in the direction of the slope, or without magnetic orientation of the pillars (perpendicular to the tilting slope, figure 9). This sliding angle is directly proportional to the contact angle hysteresis[27]. Thus, any variation of the roll-off angle means that the hysteresis has been altered. Since the advancing contact angle is almost constant whatever the surface fraction in the Cassie state, the alteration of the sliding angle may be due to a variation of receding contact angle.

In figure 9, we observed that the roll-off angle decreases when pillars are oriented against the tilting angle. When pillars are oriented in the direction of the tilting direction, the roll-off angle increases compared to the roll-off angle without magnetic actuation. In a similar experiment, Drotlef et al. evoked that the roll-off angle is increased when pillars are deflected in the direction of the tilting angle [7]. Even though not enough details are given in their articles to compare with our data, the roll-off angle they measured are larger than ours, maybe coming from a larger surface fraction than our pillar arrays. These results have not been reported elsewhere in the literature, and the present work and the apparently contradictory results from Drotlef et al. are the first to describe such a result. This will be elucidated by a more global study, for instance by varying the surface fraction of samples.

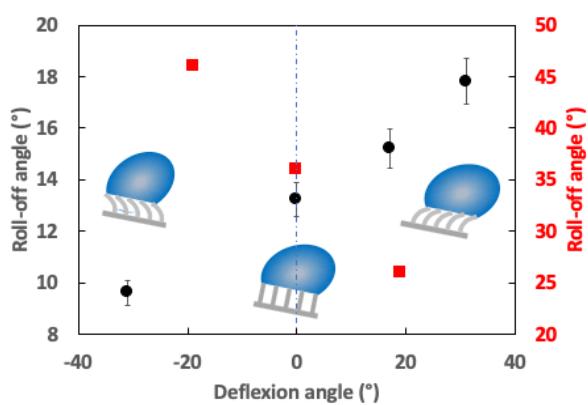

Figure 9: Roll-off angle versus deflection angle of the pillars. Black dots are our data (left scale); Red squares are data from Drotlef *et al.*[7] (right scale)

To confirm these observations, we have looked at the evaporation of a droplet. We have observed the contact line movements under an optical microscope, first with a static deflexion of the pillars, then with their periodic activation.

**Surface anisotropy and evaporation**

We followed the evolution of the droplet evaporation by looking at the contact area and its center position versus time. These experiments have been conducted in absence or in the presence of a static magnetic orientation of the pillars.

Without any magnetic orientation, the center of the contact area of the droplet remained immobile (figure 10-left). On the contrary, the position of the center of the contact area moved in the opposite direction of the direction of the pillars, thus confirming our first macroscopic observations on tilted surfaces (figure 10-center and right).

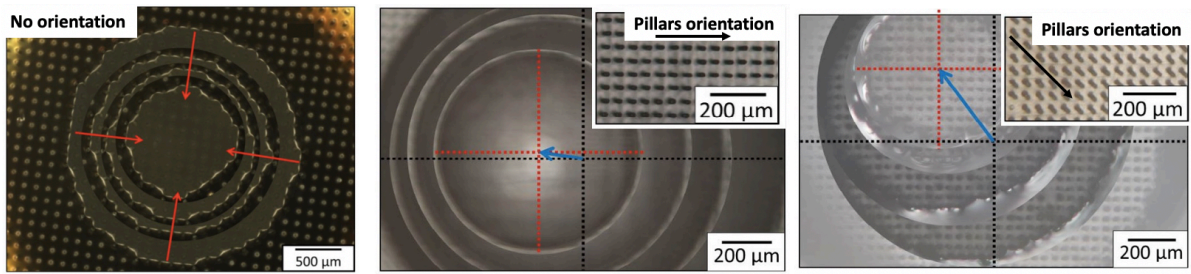

Figure 10: superposition of droplet images during evaporation without magnetic field applied (left), with pillars orientated to the right (center), and with pillars orientated down right (right)

To explain what occurs when pillars are oriented toward the center of the droplet or in the opposite direction, we hypothesize that the local receding contact angle is reached more easily when the pillars are oriented in the direction of the contact line. On the contrary, on the other side of the droplet, the contact angle value, at the level of the pillars, is comprised between the two extremum contact angle values (namely the advancing and receding ones), thus allowing a stronger anchoring of the contact line on the pillars (figure 11). The present observations are thus consistent with our roll-off angle measurements.

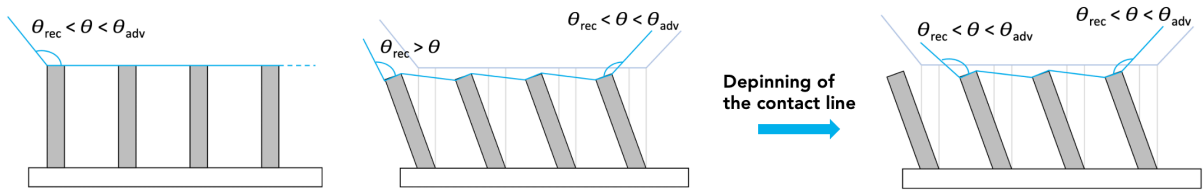

Figure 11: Scheme of the depinning process under magnetic actuation. Local contact angle becomes lower than the receding contact angle, thus the depinning of the contact line occurs.

**Asymmetric depinning during evaporation**

To highlight the role of this local receding contact angle, we followed the movement of the contact line using optical microscopy, when pillars are periodically oriented in the direction or in the opposite direction of the center of the droplet. On figure 12, and in the movie on SI-10, it is possible to see the periodical orientation of the pillars. Images A, B, C, and D, in figure 12 represent different snapshots during the dewetting induced by the movement of the pillars, up to 3 complete cycles. Figure 12-E shows the number of depinning events versus time. At the first steps, depinning is observed without pillar actuation (corresponding to the 4 pillars on the left). Then, depinning is observed mainly when pillars are orientated in the opposite direction to the center of the droplet, that is in the direction of the contact line. This is highlighted in figure 12-E with the gray areas. Moreover, the magnetic actuation induced the depinning of the contact line where no such dewetting is observed in such a short duration with classical dewetting due to evaporation. This experiment is a clear evidence that the dynamic movement of the pillar can induce the depinning of the contact line more rapidly than classical evaporation of the droplet. Additionally, the dewetting occurs mainly when the pillars are oriented in the direction of the contact line. After 8 seconds of oscillations, no more depinning is observed, mainly due to the fact that the tension on the contact line is reduced compared to its former position.

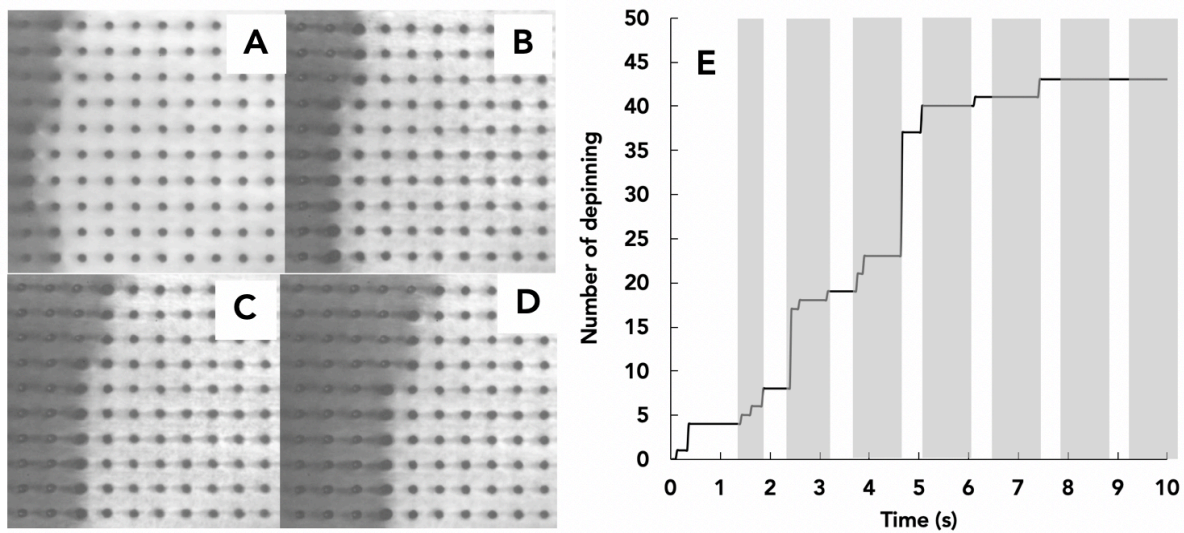

Figure 12: contact area of water (bright area) and pillars out of the droplet (darker area) at the beginning of the experiment (A), after one (B), two (C), and three (D), oscillations of the pillars (distance between two pillars: 50 µm); Pillar depinning events versus the number of pillar oscillation cycles (E). Grey areas correspond to the pillars orientated against contact line (left on the images), the other parts correspond to the pillars orientated toward the center of the droplet.

**Digital droplet dynamics**

At last, we take advantage of this local effect at the scale of the contact line and use magnetic periodic activation to trigger droplet motion at the macroscale. In previous works in the literature, a complementary force is always added to the orientation of the pillars. Seo et al. deformed the soft surface to guide a droplet along a determined trajectory [28]. Al-Azawi et al. used a magnetic field on a ferrofluid droplet to confine the droplet closed to the magnet, and extracted from these experiments the viscous damping of the droplet on the surface [29]. We here considered a surface that exhibits a sliding angle value of 13° (10 µL droplet), and measured the displacement of the droplet as the surface is tilted to a lower angle value (11°). The movement of the droplet is achieved only if the magnetic periodic movement of the magnet is applied (figure 13 and video SI-10).

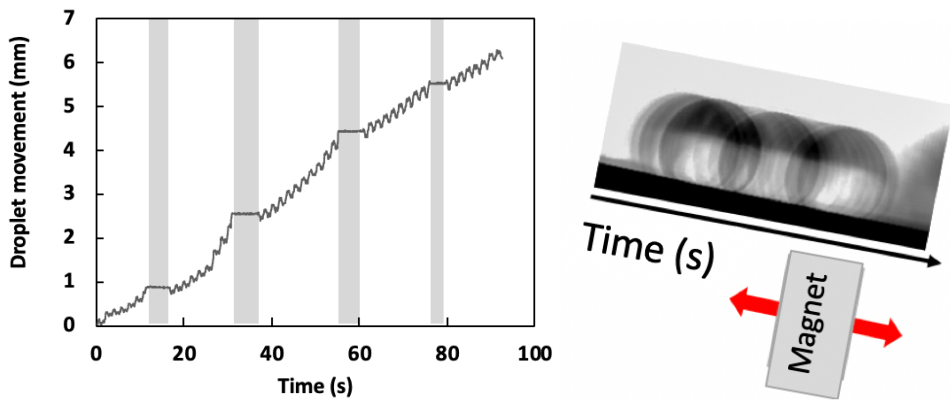

Figure 13: displacement of a droplet on magnetic pillared surface C with tilting angle of 11° (left) and 13° (right). Gray area corresponds to no-magnetic actuation.

When the magnet is stopped, the droplet immobilized. When the magnet is periodically displaced, the droplet moves back and forth as seen from the small oscillations during the global droplet displacement on figure 13. These oscillations are at the frequency of the movement of the magnet below the surface. The droplet is driven against the slope until depinning of the contact line occurs and the droplet relaxes to the new pinning position on the next row of pillars. This is in good agreement with the model presented in figure 11, where the contact line is displaced until the contact angle decreases below the receding contact angle. This is the first time that such a movement is observed, in the Cassie state, on a magnetic superhydrophobic surface. This experiment is also a good proof that an external force is mostly needed to induce some anisotropy in the movement, as depict by the results obtained by Li et al., where vibrations are needed to move a droplet in a preferential direction onto oriented pillar arrays [30]. This is in essence the limitation of this kind of activable surface. However, using a gradient of orientation of the pillars (Marangoni effect) with low surface fraction or an asymmetric movement of the pillar could help to limit the use vibration or tilted surfaces. Alternatively, the preparation of magnetic slippery surface could decrease drastically the contact angle hysteresis and help direct droplets on such surfaces [31].

## Conclusion

In this article, we prepared and characterized magnetic PDMS that can be shaped into pillar arrays with various surface fractions and aspect ratio values. We determined that the surface fraction of pillars do have an impact on the deflection of the pillars, mainly due to dipolar magnetic interactions between the pillars. We calculated the magnetic forces exerted on magnetic PDMS when the particle loading increases, and quantitatively demonstrated that this force is balanced by the elastic force. We highlighted that the magnetic dipolar interactions must be taken into account, within the described configuration of actuation, to avoid any detrimental decrease of the activation of magnetic pillars.

We explored the role of magnetic actuation on the dynamics of the contact line. When pillars are statically oriented in a direction, either the evaporation of a droplet or its sliding on a tilted surface highlighted that the droplet moves in the direction opposite to the orientation of the pillars. When pillars are periodically oriented below the surface on a tilted surface, the drops either start to move or go down the surface step by step, evidencing the role of the local anchoring of the contact line on the pillars on the macroscopic movement of the droplet. These results are highly important for the community working on magnetic pillar arrays, because they rationalize the effect of magnetic interactions between magnetic pillars, the impact of their orientation on the wetting of droplets, and the impact on the controlled macroscopic displacement of droplets on a tilted surface. In general, these magnetically activated surfaces could have high potential development for dust recovery, but they need to be particularly well defined to get the best balance between magnetic actuation and wettability.

## Supplementary informations

Movies S1-S9 correspond to pillar deflexion.

Movie S10 corresponds to the droplet displacement upon periodic activation of the pillars.

Movie SI-11 shows dynamic dipolar interactions between pillars (Surface fraction 20%)

Fundings:

This work has been supported by ANR program MADNESS .

TOC image :

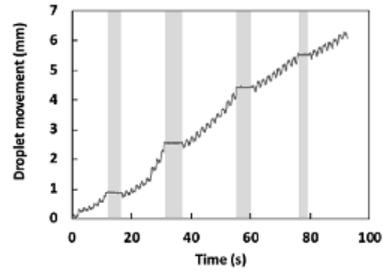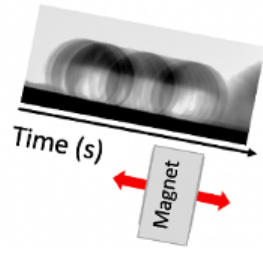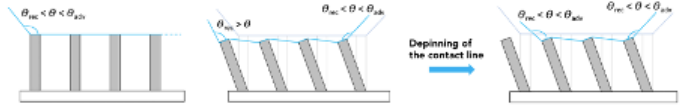